\documentclass[aps,prb,twocolumn,superscriptaddress,showpacs,floatfix]{revtex4-1}

\usepackage{amsmath}
\usepackage{graphicx}
\usepackage{color}
\usepackage{natbib,hyperref}
\usepackage{xfrac}
\usepackage{wasysym}
\usepackage{multirow}
\usepackage[caption=false]{subfig}
\usepackage{hyperref}

\hypersetup{
    linkcolor = black,
    citecolor = black,
    urlcolor  = cyan,
    colorlinks = true
}

\usepackage{xcolor,colortbl}
\usepackage[normalem]{ulem}

\definecolor{Gray}{gray}{0.85}
\definecolor{lblue}{rgb}{0.8,0.8,1}  
\definecolor{blue}{rgb}{0.6,0.6,0.9}  

\definecolor{blau}{rgb}{0.,0.,1.}
\definecolor{gruen}{rgb}{0.,0.5,0.0}  \definecolor{orange}{rgb}{1,0.625,0.}  \definecolor{rot}{rgb}{1,0.0,0.}

\newcommand{\svo}{SrVO$_3$}
\newcommand{\sto}{SrTiO$_3$}

\newcommand{\pr}{^\prime}

\newcommand{\eref}[1]{Eq.~(\ref{#1})}
\newcommand{\fref}[1]{Fig.~\ref{#1}}

\newcommand{\vek}[1]{ \hbox{\textbf #1}}
\newcommand{\svek}{\mathbf}

\newcommand{\out}[1]{}

\begin{document}
\title{
Zoology of spin and orbital fluctuations in ultrathin oxide films
}

\author{Matthias Pickem}
\email{matthias.pickem@gmail.com}
\affiliation{Institute for Solid State Physics, TU Wien,  Vienna, Austria}
\author{Josef Kaufmann}
\affiliation{Institute for Solid State Physics, TU Wien,  Vienna, Austria}
\author{Karsten Held}
\affiliation{Institute for Solid State Physics, TU Wien,  Vienna, Austria}
\author{Jan M.~Tomczak}
\affiliation{Institute for Solid State Physics, TU Wien,  Vienna, Austria}

\date{\today}

\begin{abstract}
Many metallic transition-metal oxides turn insulating when grown as films that are only a few unit-cells thick. The microscopic origins of these thickness induced metal-to-insulator transitions however remain under dispute.
Here, we simulate the extreme case of a monolayer of an inconspicuous correlated metal---the strontium vanadate SrVO$_3$---deposited on a SrTiO$_3$ substrate.
Crucially, our system can have a termination to vacuum consisting of either a SrO or a VO$_2$ top layer. While we find that both lead to Mott insulating behavior at nominal stoichiometry, the phase diagram emerging upon doping---chemically or through an applied gate voltage---is qualitatively different.
Indeed, our many-body calculations reveal a cornucopia of nonlocal fluctuations associated with (in)commensurate antiferromagnetic, ferromagnetic, as well as stripe and checkerboard orbital ordering instabilities.
Identifying that the two geometries yield crystal-field splittings of opposite signs, we elucidate the ensuing phases through the lens of the orbital degrees of freedom.
Quite generally, our work highlights that interface and surface reconstruction and the deformation or severing of coordination polyhedra in ultra-thin films drive rich properties that are radically different from the material's bulk physics.
\end{abstract}

\maketitle

\section{Introduction}
%
In strongly correlated materials\cite{RPP2017,Giustino_2021} various energy scales compete in defining the ground state.
Perturbing their balance, e.g., through pressure or doping, may induce a multitude of different
long-range orders or trigger metal-insulator transitions.\cite{imada}
Particularly rich are phenomena involving the orbital degrees of freedom.\cite{Tokura462,doi:10.1021/acs.chemrev.0c00579}
Their behavior is extremely sensitive to the local atomic environment that controls
hybridizations,\cite{jmt_radialKI} crystal-fields,\cite{Bethe1929} and their degeneracy.
Therefore, the advent of epitaxial growth of ultra-thin films and hetero-structures has
unlocked vast possibilities\cite{Tokura2003,doi:10.1146/annurev-matsci-070813-113248} to explore orbital physics and electronic anisotropies\cite{Benjamin_2D3D} in general.
Indeed, substrate strain, interfacial or surface reconstruction, and varying
surface terminations cause distortions, rotations, or even the severing of coordination polyhedra. These structural changes invariably affect the electronic, magnetic, and orbital state, often leading to properties absent in bulk samples of the same material.

\begin{figure*}
  \centering
  \subfloat[VO$_2$-terminated monolayer]{
    \includegraphics[width=0.47\textwidth]{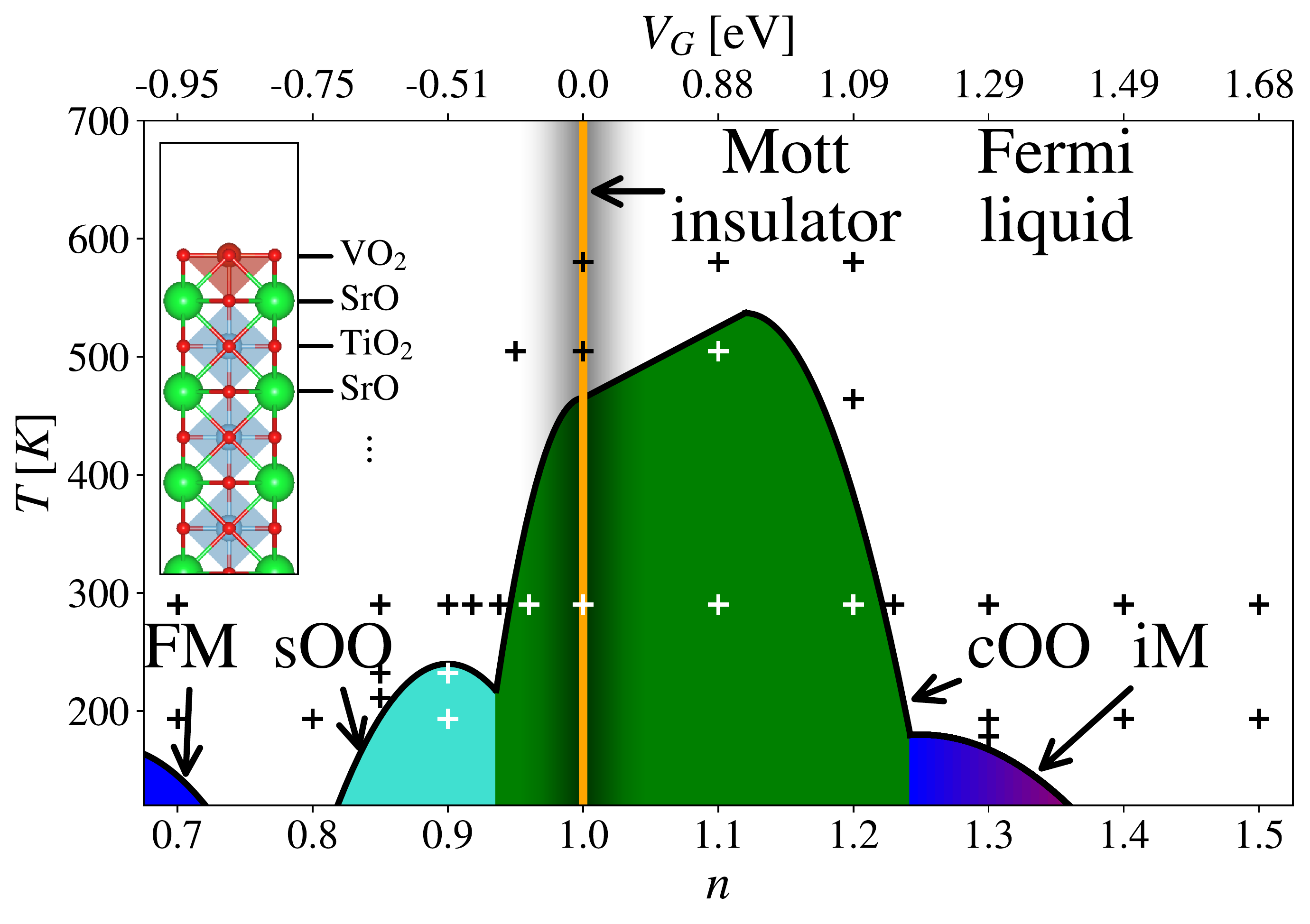}
    }
  \quad
  \subfloat[SrO-terminated monolayer]{
    \includegraphics[width=0.47\textwidth]{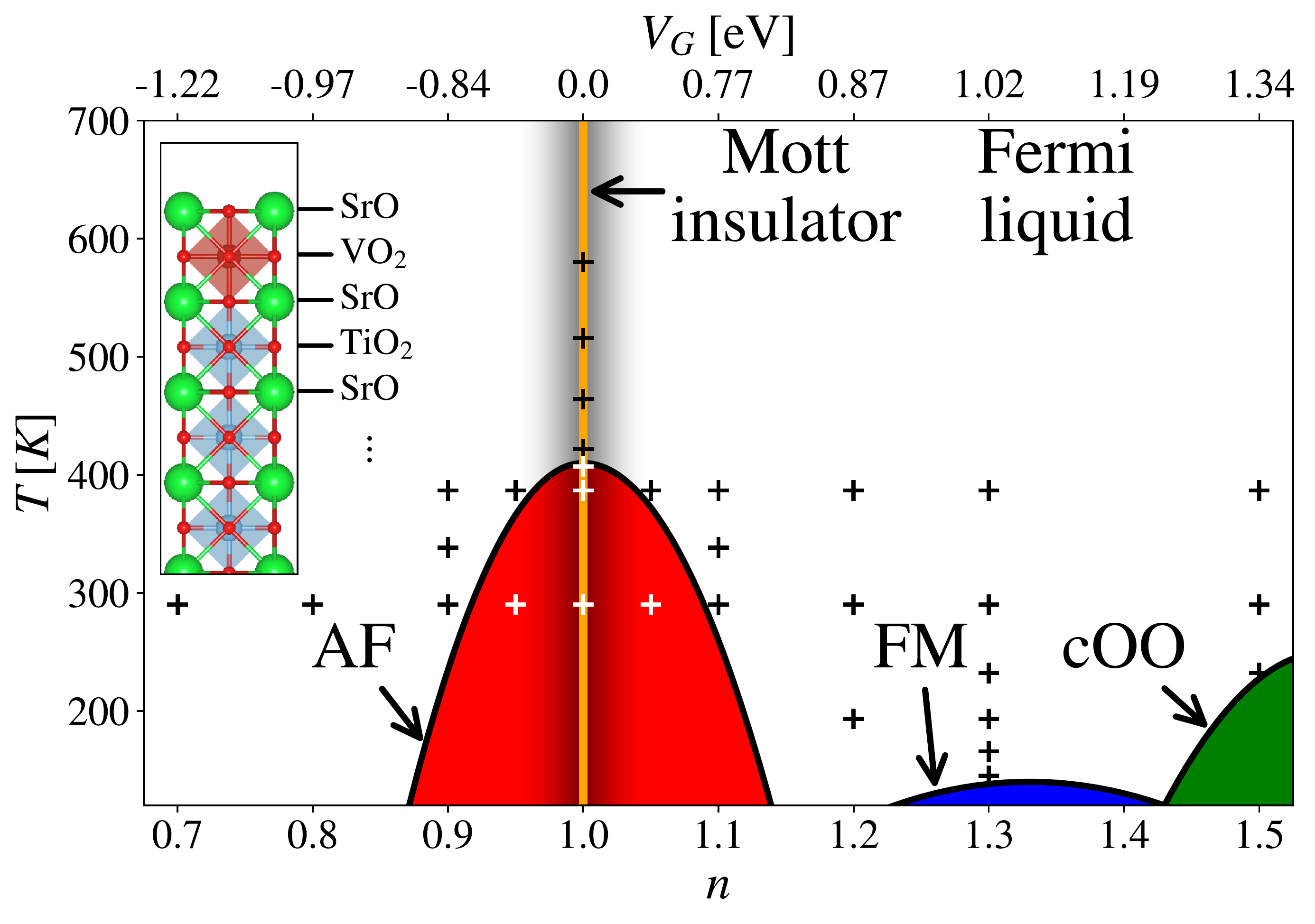}
    }
\caption{{\bf Phase diagrams.} (a) VO$_2$ and (b) SrO terminated monolayers of \svo\ on a \sto\ substrate (see insets) realize various phases as a function of the number of electrons per site in the low-energy $t_{2g}$ ($n$; lower $x$-axis) or gate voltage ($V_G$; upper $x$-axis):
    antiferromagnetism (AF: red), ferromagnetism (FM: blue), incommensurate magnetism (iM: blueish), checkerboard orbital order (cOO: green), stripe orbital order (sOO: turquoise).
    The thus colored domes indicate the formation of long-range order
    within dynamical mean-field theory (DMFT).
    The ``$+$''--signs indicate points for which many-body calculations were performed (black signs represent  non-divergent, white signs divergent DMFT susceptibilities). Based on the
    dominant susceptibilities, the color domes have been drawn as a guide to the eye.
    \label{fig:phasediagram}}
\end{figure*}

Here, we study the extreme case of a monolayer of the perovskite transition-metal oxide SrVO$_3$ deposited on a SrTiO$_3$ substrate.
A moderately correlated paramagnetic metal in the bulk,\cite{PhysRevB.58.4372,PhysRevLett.93.156402} SrVO$_3$
is known to undergo a metal-to-insulator transition (MIT) in ultra-thin films on SrTiO$_3$ below a critical thickness of 2-3 unit-cells.\cite{PhysRevLett.104.147601,Kobayashi2017}
This Mott insulator\cite{PhysRevLett.104.147601,PhysRevLett.114.246401} has further been suggested\cite{PhysRevLett.114.246401} as active material in a Mott transistor,\cite{Motttransistor} where a gate voltage is used to switch between the insulator (OFF) and a metallic (ON) state.
Besides the characterization of the charge state, however, little is known about ordered phases or long-range fluctuations in ultra-thin films. Bulk Mott insulators typically order antiferromagnetically (AF) at low enough temperatures, while doping them may lead to various types of fluctuations and symmetry-broken phases, e.g., superconductivity or charge-order.
Here, we find that the SrVO$_3$ monolayer
realizes more than five distinct phases, see \fref{fig:phasediagram}, including (in)commensurate antiferromagnetism (AF), ferromagnetism (FM), as well as stripe (s) and checkerboard (c) orbital-order (OO).
Some of these regimes can be explored by chemical doping or an applied gate voltage. Additionally, we evidence a qualitatively different phase diagram for the two possible choices of surface termination, in which the top is formed by either a VO$_2$, or a SrO layer (\fref{fig:phasediagram}a, \ref{fig:phasediagram}b, respectively).
Ultimately, we link the character of dominant fluctuations to the orbital degrees of freedom, that are tuned through the (total) filling $n$ and the crystal-field splitting $\Delta_{\rm cfs}$ between the two-fold degenerate $d_{xz}$, $d_{yz}$ orbitals and the $d_{xy}$ orbital residing in the film's plane.
With these key ingredients being common to a multitude of correlated oxides, our study of the SrVO$_3$ monolayer anticipates rich phase diagrams in transition-metal oxide ultra-thin films.

The outline of the paper is as follows: Section
\ref{Sec:Method} briefly discusses the methods employed: density functional theory (DFT) and dynamical mean-field theory (DMFT).
Section \ref{sec:DFT} presents the quantitative differences between the two different surface terminations on the one-particle level.
Section \ref{sec:Mott} presents a many-body discussion of the Mott insulating state found in the stoichiometric systems, while Section \ref{Sec:DMFT} presents the electronic structure trends upon doping the SrVO$_3$ monolayers.
Section \ref{Sec:DMFTsusc} characterizes magnetic and orbital fluctuations
and associated ordering instabilities on the basis of DMFT susceptibilities.
Section \ref{Sec:Discussion} puts our findings into perspective.
Finally, Section \ref{Sec:Conclusion} summarizes the results and provides an outlook.

\section{Method}%
\label{Sec:Method}
DFT calculations are performed with the WIEN2k package \cite{wien2k,doi:10.1063/1.5143061}
using the PBE\cite{PhysRevLett.77.3865} exchange-correlation potential.
We construct systems with one half of a unit cell of \svo~(VO$_2$ termination) and one unit cell of \svo~(SrO termination) on top of a substrate of six unit cells of \sto\ surrounded by sufficient vacuum of around $10$\AA\ in z-direction 
(see insets in Fig.~\ref{fig:phasediagram}). In both setups the transition between \sto~and \svo~consists of a TiO$_2$ - SrO - VO$_2$ interface, consistent with experiment \cite{C6CP07691B,PhysRevB.100.155114}, while at the bottom the \sto~substrate is terminated via SrO to vacuum.
Since experimentally \svo\ is locked to the \sto\ substrate\cite{PhysRevMaterials.3.115001}
we initialize the in-plane lattice constant
with the PBE-optimized value for bulk \sto\
$a_{\mathrm{SrTiO}_3}=3.95$\AA.
To treat the surface properly the two unit cells of \sto\ furthest away from \svo\ are then constrained to $a_{\mathrm{SrTiO}_3}$, simulating the transition to the SrTiO$_3$ bulk, while all other internal atomic positions are fully relaxed.

\begin{figure*}[!t]
  \centering
  \subfloat[VO$_2$-terminated monolayer: $\Delta_{\rm  cfs}<0$]{
    \includegraphics[width=0.47\textwidth]{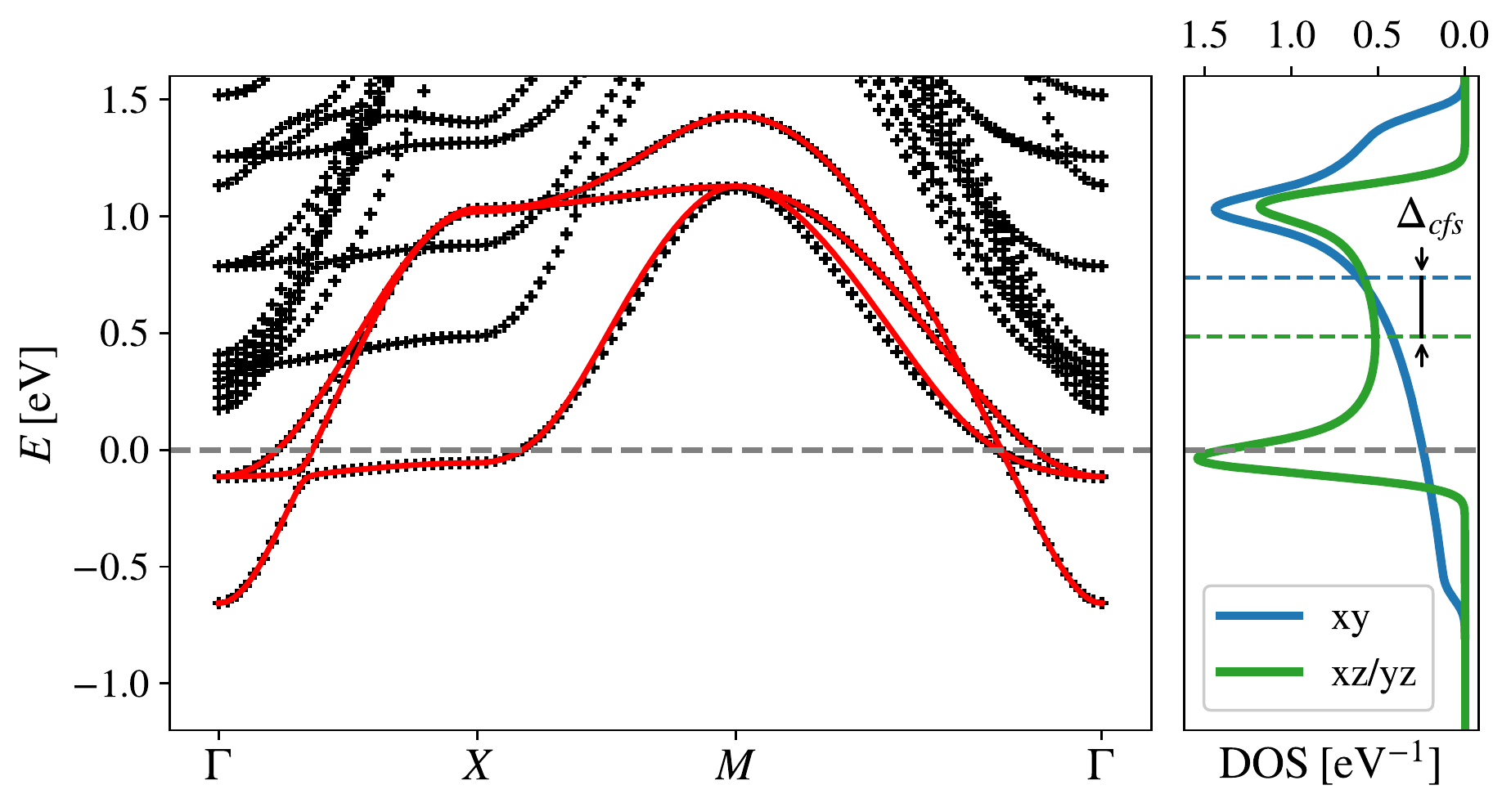}
    }
    \quad
  \subfloat[SrO-terminated monolayer: $\Delta_{\rm cfs}>0$]{
    \includegraphics[width=0.47\textwidth]{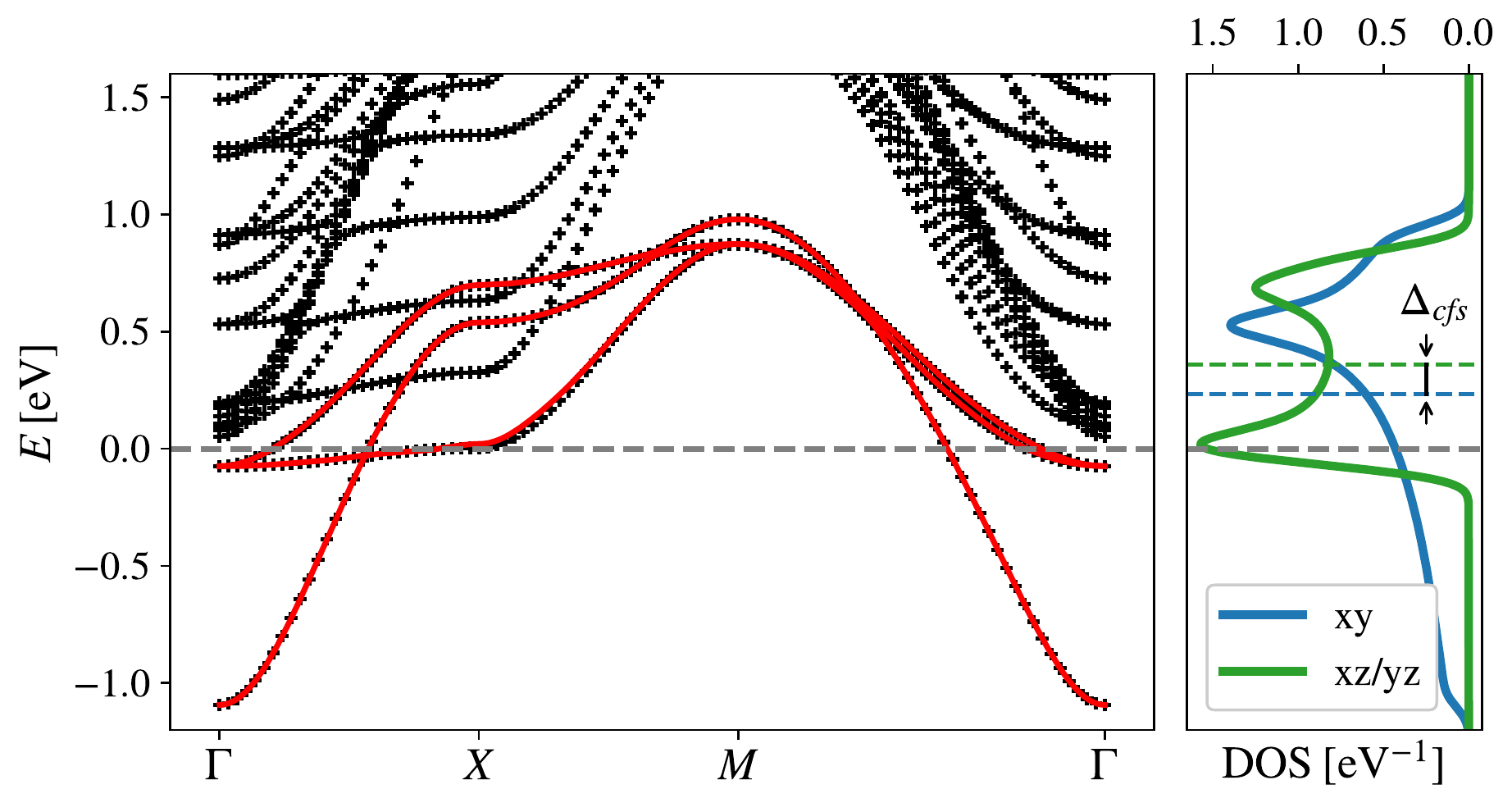}
    }
  \caption{{\bf Band structures and density of states.} (a) VO$_2$ and (b) SrO terminated monolayer. Left: DFT band structure along a momentum path through the Brillouin zone (black dots) overlain with the
  t$_{\mathrm{2g}}$-orbital projected Wannier dispersion (red lines). Right: the resulting
  density of states (DOS) clearly shows the (quasi--) 2D character of the $xy$-orbital
  (blue) and the 1D character of the locally degenerate $xz/yz$-orbitals (green). The local orbital energy levels are marked as dashed horizontal lines in the DOS.
  We find a crystal field splitting of $\Delta_{\rm cfs} = -0.252$eV for the VO$_2$ terminated layer and $\Delta_{\rm cfs} = +0.126$eV for the SrO terminated monolayer.}
\label{Fig:dft}
\end{figure*}

We then perform Wannier projections
onto maximally localized V-t$_{\mathrm{2g}}$ orbitals, using the WIEN2Wannier \cite{wien2wannier} interface to Wannier90 \cite{wannier90}, see Fig.~\ref{Fig:dft}.
These Wannier
Hamiltonians are supplemented with an effective SU(2)-symmetric Kanamori
interaction of $U=5$eV, $J=0.75$eV, $U'=3.5$eV
(similar to Ref.~\onlinecite{PhysRevLett.114.246401}),
for which we perform dynamical mean-field theory (DMFT) \cite{bible,doi:10.1080/00018730701619647}
calculations at various temperatures.
For the undoped bulk, this setup yields the correct mass enhancement and spectra qualitatively congruent with photoemission spectroscopy\cite{pavarini:176403, Mo2003}. These values can be thought of as a lower boundary since in ultrathin films, interaction parameters increase slightly\cite{PhysRevLett.114.246401} with respect to their bulk values.
The Hamiltonians are kept constant under doping.
The analytic continuations to real frequencies are performed with the maximum entropy method\cite{maxent} used in the \verb=ana_cont= library\cite{PhysRevLett.122.127601,kaufmann2021anacont}.

The DMFT self-consistency cycle as well as the sampling of the two-particle Green's function is done by continuous-time quantum Monte Carlo simulations in the hybridization expansion \cite{Werner2006,Gull2011a} using w2dynamics \cite{w2dynamics} with worm sampling \cite{Gunacker15}.
Momentum-dependent DMFT susceptibilities are calculated from the local vertex,
using the  AbinitioD$\Gamma$A \cite{Anna_ADGA,CPC_ADGA} program package.

\section{Results}
\subsection{Surface termination, band structure and density of states}
\label{sec:DFT}
Both, SrO and VO$_2$-terminated films result in similarly looking DFT band structures whose relevant orbitals around the Fermi level are of vanadium $t_{2g}$ character, see Fig.~\ref{Fig:dft}.
The densities of states (DOS) of these orbitals showcase the abrupt surface termination of the sample: While the $xy$-projection (blue line) keeps its two-dimensional character (as in bulk SrVO$_3$) the (locally degenerate) $xz$- and $yz$-projections (green lines) now become one-dimensional and, concomitantly, display a strongly reduced bandwidth. Consequently a van-Hove singularity emerges, which, at zero doping, is in close proximity to the Fermi level. Indeed, in the VO$_2$ (SrO) terminated system this singularity is situated slightly below (above) the Fermi level.
On top of this dimensionality reduction we find the crystal-field splitting (cfs)
\begin{equation}
    \Delta_{\rm cfs} = E_{xz/yz}-E_{xy},
\end{equation}
to have opposite signs for the two different setups:
The {\it positive} cfs of the SrO terminated monolayer, $\Delta_{\rm cfs}=+0.13$eV, is a direct result
of the tensile strain caused by the (in-plane) lattice mismatch between SrVO$_3$ and SrTiO$_3$ ($a_{\mathrm{SrVO}_3} < a_{\mathrm{SrTiO}_3}$).
The in-plane expansion triggers a structural compression in the perpendicular direction, so as to keep the volume approximately constant.
This structural anisotropy translates into an electronic anisotropy\cite{Benjamin_2D3D}:
The evident breaking of the cubic symmetry of SrVO$_3$ lifts the three-fold $t_{2g}$ degeneracy, making the $xy$-orbital energetically favorable.
The same effects take place in the VO$_2$ terminated monolayer as well. There, however, the geometric distortion gets overcompensated by the missing SrO layer: severing the apical oxygen of the transition metal coordination octahedron results in a reversed, {\it negative} $\Delta_{\rm cfs}=-0.25$eV. The $xz/yz$ orbitals have their lobes pointing in the $z$-direction, towards the lobes of the oxygen $p_x/p_y$ orbitals. The missing overlap to the absent apical oxygen leads to less electrostatic repulsion, thus lowering the energy required to occupy these states.
Another contributing factor is the abrupt termination to vacuum, removing any restriction in the positive $z$-direction for the structural relaxation:
The VO$_2$ (SrO) terminated system results in a concave (convex) final termination, i.e., the last VO$_2$ (SrO) layer bends inwards (outwards).
For both systems we find almost identical $xy$ orbitals with a nearest-neighbor hopping $t_{xy}\sim-230$meV, next-nearest-neighbor hopping $t\pr_{xy}\sim-70$meV and bandwidth W$_{xy}\sim2.1$eV.
The $xz$ and $yz$ orbitals of both systems, on the other hand, can be described purely by nearest-neighbor hopping along the $x$ or $y$ direction, respectively: The VO$_2$-terminated monolayer allows for a large hopping amplitude ($t_{xz/yz}\sim-300$meV), resulting in a slightly larger bandwidth $W_{xz/yz}=1.2$eV in Fig.~\ref{Fig:dft}a, compared to only $W_{xz/yz}=0.95$eV ($t_{xz/yz}\sim-200$meV) for the SrO-terminated monolayer in Fig.~\ref{Fig:dft}b.

Recent experiments\cite{PhysRevLett.119.086801,Gabel2021} suggest 
ultra-thin SrVO$_3$ films to be VO$_2$-terminated. However, for reasons of stability (surface oxidization, surface protection, etc.), a SrO-boundary could be preferable. This might be achievable either implicitly via a SrTiO$_3$ capping-layer (see Sec.~\ref{Sec:Conclusion}) or explicitly via deposition of SrO on top of the VO$_2$ surface. The latter has in fact been achieved 
for LaNiO$_3$ films on LaAlO$_3$ substrate, where both LaO and NiO$_2$ terminations are possible through an ablation of La$_2$O$_3$ and NiO, respectively\cite{Golalikhani2018}.

\subsection{Stoichiometric Mott insulator}
\label{sec:Mott}
\begin{figure}[!t!h]
  \centering
  \includegraphics[width=0.48\textwidth]{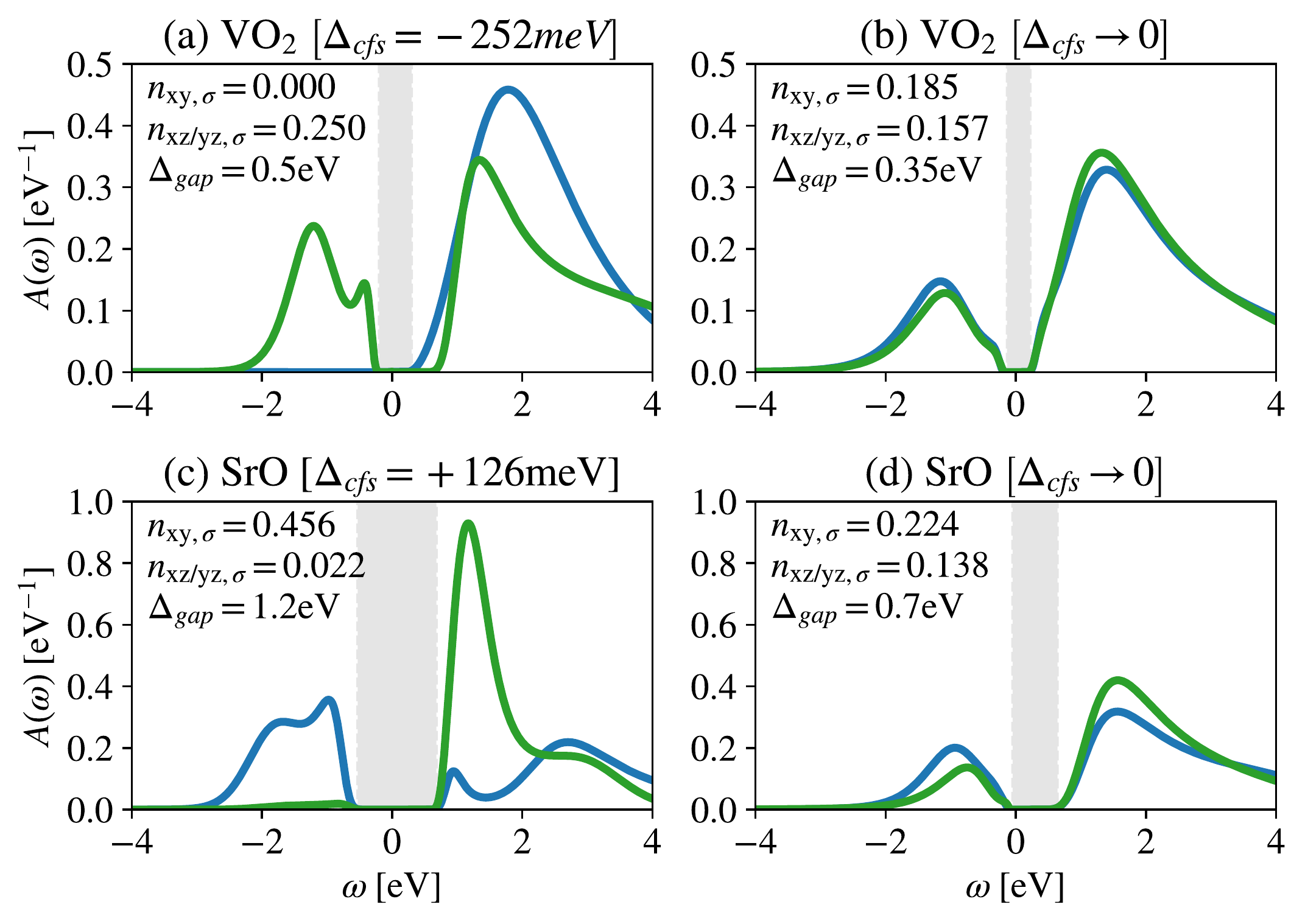}
  \caption{{\bf Mott insulating ground state.} DMFT spectral functions $A(\omega)$ for (a) the VO$_2$ terminated and (c) the SrO terminated structure at $U=5$eV and room temperature ($T=290$K). In both cases a wide Mott gap forms which is accompanied by a strong orbital polarization.
  Removing the crystal-field splitting, the reduced bandwidth alone results in a  Mott insulator (b,d) with a slightly smaller band gap.
  }
 \label{fig:mott}
\end{figure}

Next, we analyze the electronic structure for the stoichiometric samples ($n=1$)
on the many-body level, using DMFT at room temperature $T=290$K:
While in the VO$_2$-terminated monolayer (\fref{fig:mott}a) the out-of-plane $xz/yz$-orbitals realize a {\it quarter-filled} Mott insulator with a gap of $0.5$eV, in the SrO terminated monolayer (\fref{fig:mott}c) the in-plane ${xy}$-orbital hosts an essentially {\it half-filled} canonical Mott insulator with a gap of $1.2$eV.
This difference can be traced back to the bare crystal-fields.
Indeed, DMFT amplifies the effect of the DFT cfs for both terminations, leading to the depletion of the energetically higher lying orbital(s), i.e., the $xy$ and the $xz/yz$ orbital(s) for the VO$_2$ and SrO termination, respectively.
This correlation-enhanced orbital polarization\cite{poter_v2o3} leads to an effectively reduced orbital-degeneracy. As a consequence, charge (inter-orbital) fluctuations are suppressed and the critical interaction for reaching the Mott state diminishes \cite{PhysRevB.54.R11026,PhysRevB.70.205116,pavarini:176403}: The Coulomb interaction is large enough to open a Mott gap in the SrO (VO$_2$) terminated monolayer with a single (two-fold) degenerate lowest orbital, while three-fold orbitally degenerate bulk SrVO$_3$ is a stable metal.
Let us note that the evidenced orbital polarization persists when including charge self-consistency, which only yields minor corrections because charge is only redistributed between orbitals, not between sites\cite{PhysRevB.94.155131}.

However, for both terminations the insulating behavior is actually driven by a {\it combination} of
the crystal-field splitting\cite{PhysRevLett.114.246401}
and the reduced band-widths\cite{PhysRevLett.104.147601}.
Whereas the crystal-field splitting is essential for the bilayer system \cite{PhysRevLett.114.246401}, we find that the bandwidth reduction alone is sufficient to drive the monolayers insulating.
We illustrate this in \fref{fig:mott}b and \fref{fig:mott}d where we take the original Hamiltonians and set the cfs artificially to zero by shifting the local orbital energies. Both systems remain insulating in DMFT. Unswayed by the cfs, however, the Mott gaps turn out smaller and orbital occupations (in both cases: $n_{\mathrm{xy},\sigma}>n_{\mathrm{xz/yz},\sigma}$) only reflect the asymmetry of the orbitals' DOS. Let us note here that if we instead keep the cfs unchanged and adjust the $xz/yz$-bandwidths such that $W_{xz/yz} = W_{xy}$ both systems remain firmly metallic.

\begin{figure}[!t!h]
  \centering
  \includegraphics[width=0.46\textwidth]{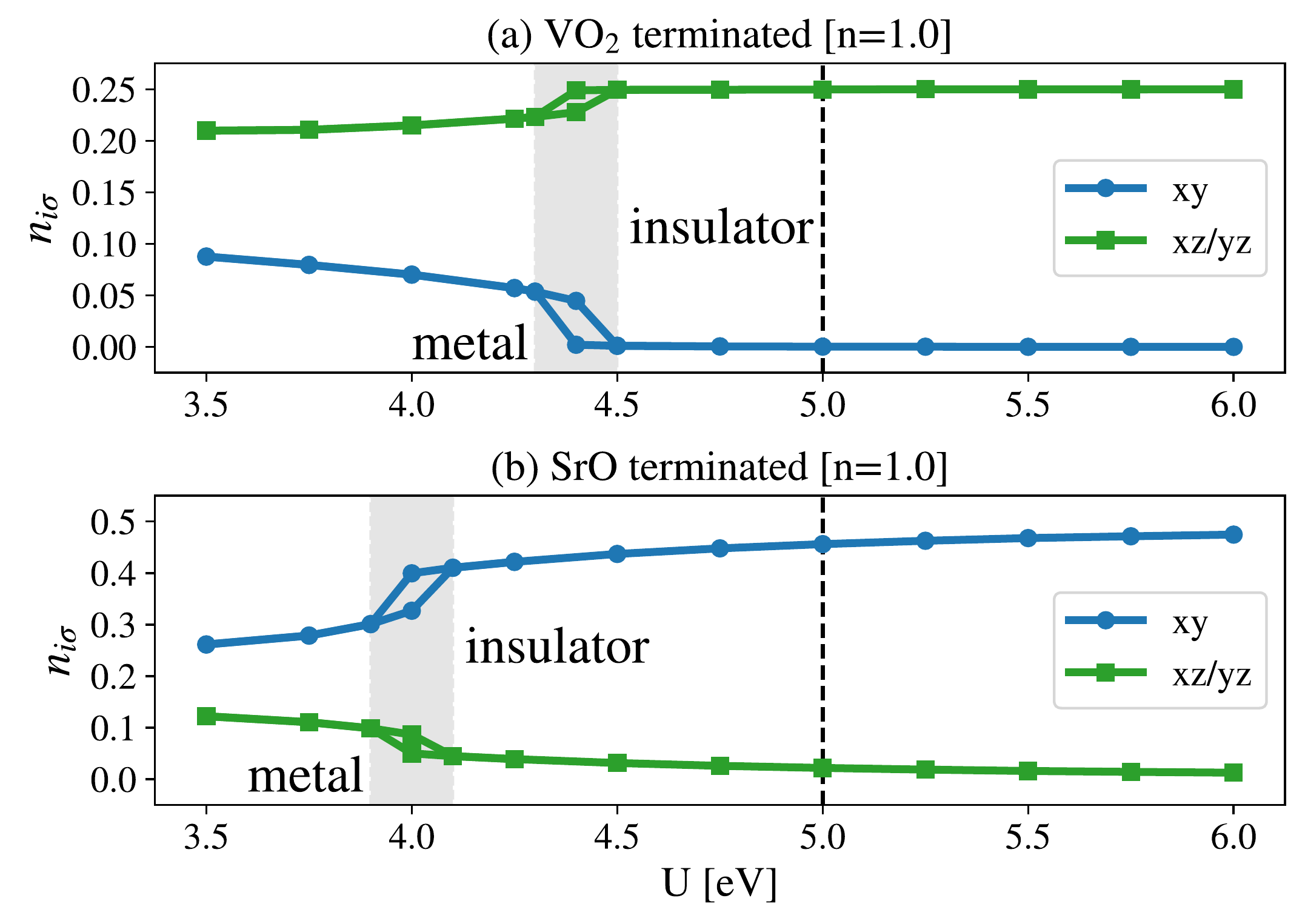}
  \caption{{\bf Stability of Mott insulating state.}
   Spin-dependent orbital occupation $n_{i\sigma}$ vs.~intra-band interaction strength $U$ at room temperature $T=290$K; Hund's coupling  $J=0.75$eV; inter-band interaction $U\pr = U-2J$. (a) The VO$_2$ terminated monolayer is effectively a two-orbital quarter-filled system ($n_{xz/yz,\sigma}\sim 0.25$) while (b) the SrO terminated monolayer becomes effectively a half-filled one-orbital system ($n_{xy,\sigma}\sim 0.5$) at large enough interaction strengths. Both lead to a Mott localization of carriers which can be upheld even if we reduce the interaction. The transition to the metallic solution is accompanied by a tight hysteresis after which the orbital polarization drops rapidly. The calculations under doping in the next figure are performed for $U=5$eV (vertical, dashed black line).}
 \label{fig:stability}
\end{figure}

To investigate the stability of the insulating state further, we perform DMFT calculations for various intra-orbital interaction strengths $U$. While keeping the Hund's coupling $J$ fixed to $0.75$eV,\footnote{Note that, in contrast to $U$, $J$ is hardly screened so that there is much less uncertainty and ambiguity than for $U$.}  we adjust the inter-orbital interaction strength $U\pr$ according to spherical symmetry ($U\pr=U-2J$)\cite{PhysRevB.90.165105}.
Fig.~\ref{fig:stability} shows the orbital occupations depending on the interaction $U$:
Starting from our standard value $U=5$eV (vertical dashed line), going to larger interaction strengths simply stabilizes the insulating solution further, while also increasing the orbital polarization slightly. Smaller interaction strengths on the other hand, reduce the orbital polarization until, eventually, the insulating solution can no longer be stabilized.
This metal-to-insulator transition is, as expected within DMFT, of first order (hysteresis or coexistence regime marked in gray in \fref{fig:stability}) and manifests itself by a sudden drop of the orbital polarization.
The Mott insulating state is stable down to $U=4.5$eV ($U=4.1$eV) for the VO$_2$ (SrO) terminated monolayer.

The stability of the stoichiometric Mott insulating solution is in particular important when doping away from it, see Sec.~\ref{Sec:DMFT}. As long as the stoichiometric sample is insulating, we expect that any
variation of the interaction will have no qualitative impact on the DMFT phase diagram. A smaller on-site repulsion will merely lead to weaker orbital polarizations and shifted boundaries in the phase diagrams,  \fref{fig:phasediagram}a,b.

On top of the Mott physics discussed here, weak localization through disorder may play an additional role in the insulating behavior of transport properties\cite{PhysRevB.100.155114}.
However, the suppression of the one-particle spectra for thin films\cite{PhysRevLett.104.147601}, magneto-transport results for SrVO$_3$ thin films on an LSAT substrate\cite{https://doi.org/10.1002/admi.201300126} as well as SrVO$_3$/SrTiO$_3$ supperlattices\cite{Wang2020} argue against a dominant weak localization scenario for the insulator. Similar observations have been made for CaVO$_3$ thin films on SrTiO$_3$ substrate\cite{doi:10.1063/1.4798963}.

\subsection{Behavior under doping}
\label{Sec:DMFT}

\begin{figure}[!t]
  \includegraphics[width=0.47\textwidth]{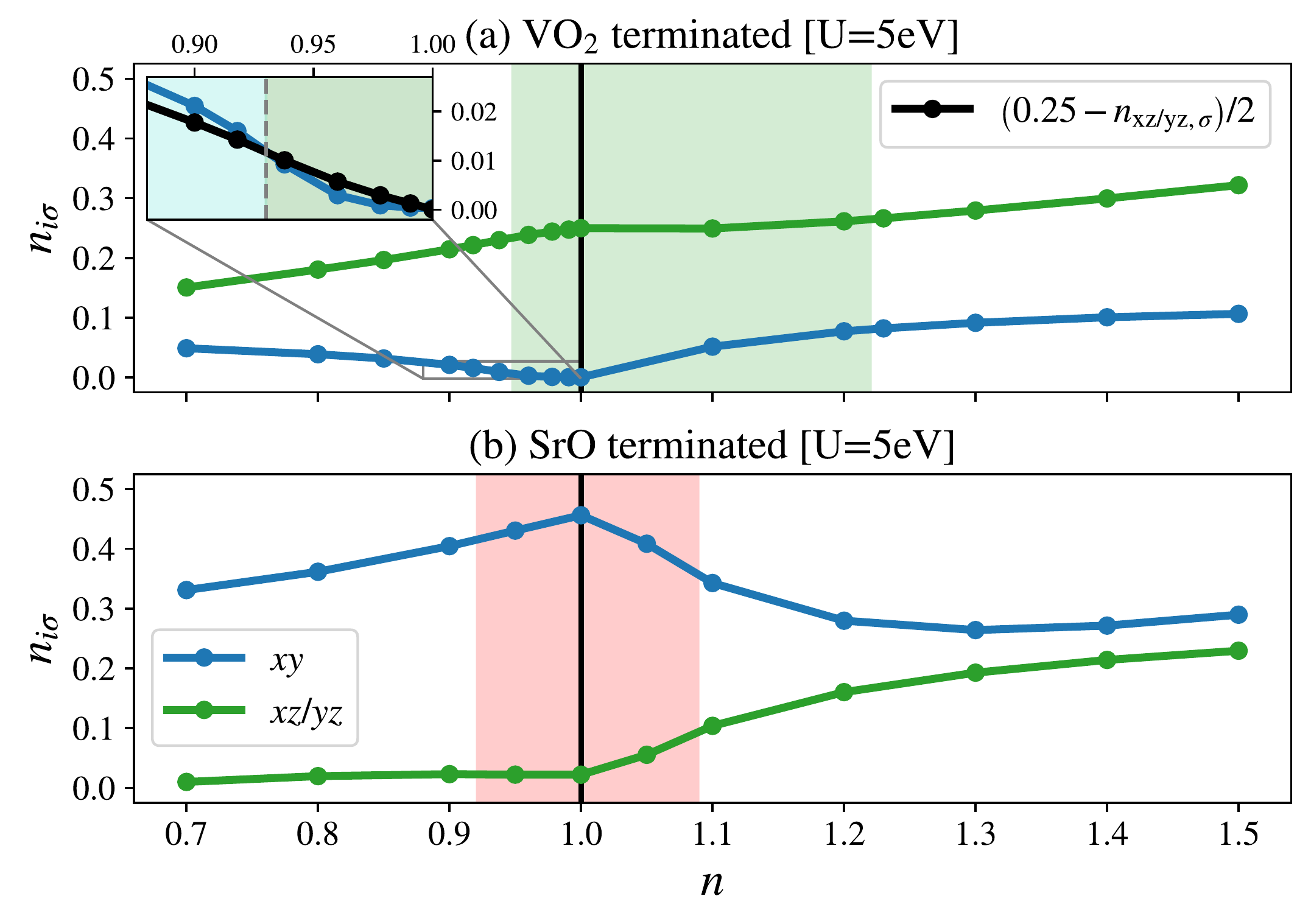}
  \caption{{\bf System trends with doping.} Spin-dependent orbital occupation at $T=290$K, resolved for the $i=xy$ and $xz$/$yz$-orbitals.
  (a) The VO$_2$ terminated system around nominal filling ($n = 1$)
  can be effectively described by two quarter-filled (xz/yz) orbitals.
  (b) The SrO terminated system at and below nominal filling can effectively be described by a single half-filled (xy) orbital.
  The former explains the tendency for checkerboard orbital ordering, while the latter promotes antiferromagnetism in Fig.~\ref{fig:phasediagram}.
  The shaded areas around undoped \svo\ represent the doping levels
  at which we find these checkerboard and antiferromagnetic orderings in DMFT at this temperature.
  Inset: Semi-empirical condition where stripe orbital ordering emerges: If we frustrate the local site enough, we find a transition from checkerboard to stripe orbital-order (indicated by a change in background color)}
  \label{fig:occupation}
\end{figure}


We now discuss the electronic structure of the doped monolayers in their
non-symmetry broken phases. From the information of orbital occupations and degeneracies, we motivate possible ordering instabilities (that will then be quantitatively assessed in Sec. \ref{Sec:DMFTsusc}).
First, the stoichiometric insulating state in \fref{fig:mott} and the various orbital occupations in \fref{fig:occupation}, indicate that
our systems are asymmetrical with respect to doping with electrons ($n>1$) or holes ($n<1$).
The VO$_2$ terminated monolayer (\fref{fig:occupation}a) somewhat upholds its orbital polarization when holes are introduced to the system. Such orbital occupations of the bipartite lattice
system make the system prone to a staggered, checkerboard orbital ordering (cOO) \cite{Held1998,PhysRevB.58.R567}, as two orbitals may now be occupied alternately on neighboring lattice sites. This is energetically favorable, since a nearest-neighbor hopping then results in a state where different orbitals are occupied so that a (virtual) hopping process costs only $U'=U-2J$ instead of $U$.

Doping with more holes, see \fref{fig:occupation}a, the VO$_2$-terminated monolayer
quickly moves away from quarter-filling by redistributing electrons from the $xz/yz$-orbitals into the $xy$-orbital,
As we shall see in Sec.~\ref{Sec:DMFTsusc}, before OO is fully suppressed
upon hole doping the ordering vector changes to stripe orbital ordering (sOO)
above a particular filling of the so far auxiliary $xy$-orbital.
Electron doping on the other hand maintains the quarter-filled state for longer, where for fillings up to $n\sim 1.2$ the additional electrons solely occupy the $xy$-orbital. Only above $n\sim 1.3$ we see a coincidental increase of all orbital occupations,
again disfavoring orbital order.

The SrO-terminated monolayer (\fref{fig:occupation}b) is even more asymmetrical: Introducing holes does not affect the effective one-band description of the system and the sparsely filled $xz/yz$ orbitals remain almost depleted.
More interesting is the electron-doped side, where the multi-orbital
character is promoted.
In this electron-doped regime, the Hund's coupling $J$ will promote a parallel spin alignment of the electrons in the three orbitals. It is natural to expect that the hopping transfers this local spin alignment into an FM order on the lattice, but other orders such as OO may emerge here as well \cite{Tokura2000}.
For strong Coulomb interactions and in an insulting state, these competing phases can be understood by superexchange as in the classical Kugel-Khomskii spin-orbital models \cite{KHOMSKII1973763}.
These phases have also been found in early DMFT calculations for a two-band model \cite{Held1998} and an oversimplified Stoner criterion predicts FM order of
the $m$-fold degenerate Hubbard model for
$A(0)\left(U+(m-1)J\right)\geq 1$\cite{Fazekas_book}.
At extremely large dopings around $n\sim 1.5$, the physics changes once again: The system now consists of three quite equally filled orbitals where the $xz/yz$ orbitals approach quarter-filling. Similarly to stoichiometric filling in the VO$_2$ terminated monolayer, such degenerate quarter-filled orbitals may lead again to orbital ordering.
\subsection{DMFT susceptibilities}
\label{Sec:DMFTsusc}
\begin{figure}[!t]
  \includegraphics[width=0.47\textwidth]{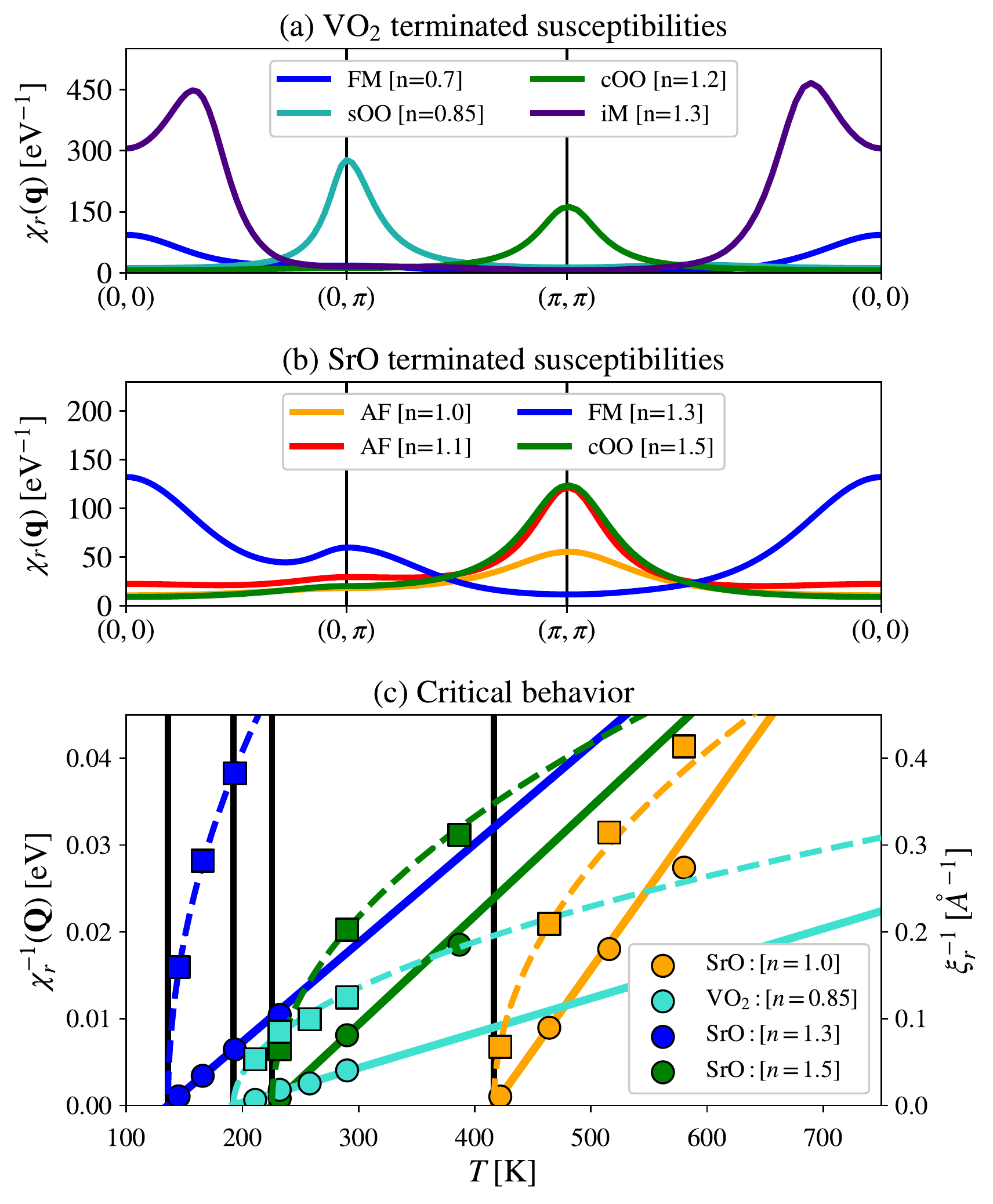}
    \caption{{\bf DMFT susceptibilities and criticality.} Momentum-dependence of the susceptibility $\chi_r(\vek{q})$ for (a) the VO$_2$ terminated and (b) the SrO terminated SrVO$_3$ monolayer in the vicinity of the respective phase transitions.
    The dominating component can be either found directly in the magnetic channel (r=`m', Eq.\eqref{eq:suscmagn}) or be obtained via a linear orbital combination of the density channel (r=`d', Eq.\eqref{eq:suscoo}).
    (c) Temperature-dependence  of the inverse DMFT  susceptibility $\chi_r^{-1}(\vek{Q})$ (first diverging $r$, $\vek{Q}$ at selected dopings; circles, left axis) for selected points from (a) and (b); lines are linear fits.
  Intersections with the $T$-axis denote the transition temperature for the respective order.
  On the secondary (right) axis the corresponding inverse correlation lengths $\xi_r^{-1}$ are shown (squares, right axis); dashed lines are fits to mean-field behavior.}
  \label{fig:suscq}
\end{figure}

\begin{figure}[!t]
  \includegraphics[width=0.48\textwidth]{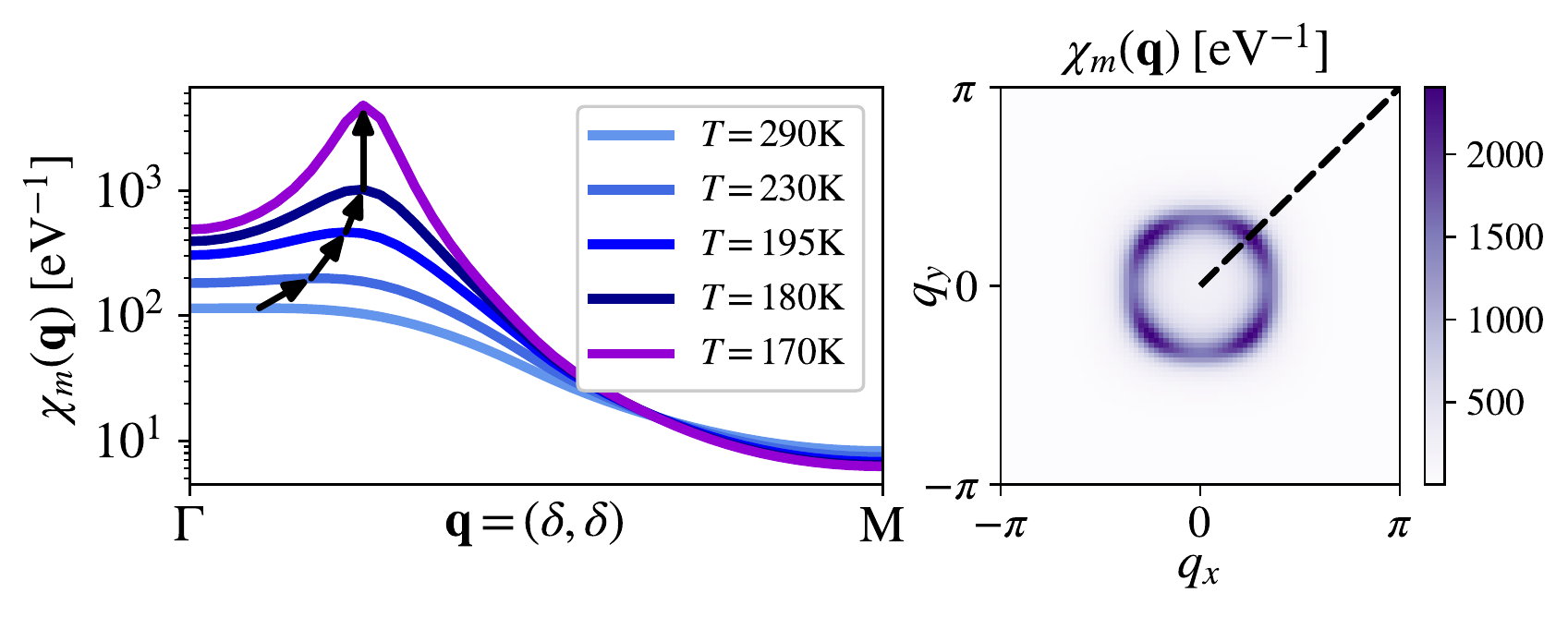}
  \caption{{\bf Incommensurate magnetism (iM).} Left: magnetic susceptibility $\chi_m(q)$ along the high-symmetry path $\Gamma\rightarrow \mathrm{M}$ for the VO$_2$-terminated monolayer at $n=1.3$ at various temperatures. Right: magnetic susceptibility $\chi_m(q)$ for $T=170$K in the planar Brillouin zone. The susceptibility is maximal roughly on a circle centered around $\mathbf{q}=(0,0)$. Upon lowering temperature the maximum of the susceptibility moves to larger $\mathbf{q}$-vectors. Close to the transition temperature we find anisotropy on this circle where the susceptibility is clearly maximal (purple)  for $\mathbf{q} = \left(\pm\delta,\pm\delta\right)$.}
  \label{fig:incomm}
\end{figure}

We now put the above analysis of potential ordered phases on firm footing:
For the prevailing magnetic and orbital orders,
\fref{fig:suscq}a and \fref{fig:suscq}b displays the relevant DMFT susceptibilities at temperatures above the respective instabilities.
Maxima in the shown susceptibilities indicate type and $\mathbf{Q}$-vector
of the dominant fluctuations.
\fref{fig:suscq}c illustrates, for selected examples, the critical behavior of the (inverse) susceptibilities and (inverse) correlation lengths
emerging when said maxima turn into instabilities.
Magnetic instabilities occur where the static susceptibility in the magnetic channel
\begin{equation}
    \chi_m(\vek{Q})
    = g^2 \sum_{ij,ll\pr}e^{i\svek{Q}(\svek{R}_i-\svek{R}_j)}\int d\tau 
    \left\langle T_\tau S^z_{il}(\tau)S^z_{jl\pr}(0)  \right\rangle
\label{eq:suscmagn}
\end{equation}
diverges at a critical temperature, indicated by the intercept of $\chi^{-1}(\mathbf{Q})$ with the temperature axis in \fref{fig:suscq}c.
In \eref{eq:suscmagn}, $g$ is the Land\'e factor; $i,j$ are indices for the lattice sites ${\mathbf R}_{i(j)}$; $l,l\pr$ are 
orbital indices. The $z$-component of the spin operator $S_{il}^z=(n_{il \uparrow}-n_{il \downarrow})/2$ is expressed in terms of the 
number operator $n_{il\sigma}$ for an electron on site $i$ in orbital $l$ with spin $\sigma$.
Ferromagnetism (FM) and antiferromagnetism (AF) correspond to the usual ordering vectors $\vek{Q}=(0,0)$ and $\vek{Q}=(\pi,\pi)$, respectively. The incommensurate magnetism (iM), that we find for the VO$_2$-termination, corresponds to an ordering vector $\vek{Q}$ with fixed length $\left|\vek{Q}\right| = \delta \geq 0$, see \fref{fig:suscq}a and \fref{fig:incomm}.

We now assess the instabilities resulting in the phase diagrams, \fref{fig:phasediagram}a,b.
{\it Antiferromagnetic} (AF) order from super-exchange is facilitated by effectively half-filled orbitals. For the $d^1$ configuration of SrVO$_3$, only the SrO-terminated monolayer provides this favorable condition. Indeed, there, the positive crystal-field
realizes a half-filled, Mott insulating $xy$-orbital that then hosts AF order,
see the diverging susceptibility in \fref{fig:suscq}(c).
Note that AF order was also predicted for a SrO-terminated SrRuO$_3$ monolayer on SrTiO$_3$ around nominal stoichiometry.\cite{Liang_SRO}
There, the $d^4$ configuration results in an essentially fully occupied $xy$-orbital, and the staggered moment is instead carried by half-filled $xz/xz$ orbitals.
In both cases, doping with either electrons or holes suppresses the AF state.

Doping the SrVO$_3$ monolayer with either termination towards their respective van-Hove singularities, i.e., hole (electron) doping for the VO$_2$ (SrO) terminated sample (see \fref{Fig:dft}) results in a strongly increased spectral density around the Fermi level 
within DMFT. 
Concomitantly doping 
generates an orbital configuration where all involved orbitals are close to equally filled, promoting energy minimization through Hund's exchange $J$ and therefore a parallel alignment of the involved spins.
This situation, leading to {\it ferromagnetism} (FM), is found around $n=0.7$ in the VO$_2$-terminated and near $n=1.3$ in the SrO-terminated sample. In both cases, ferromagnetism is hosted by the degenerate $xz$ and $yz$-orbitals.
Subleading non-local AF fluctuations are, however, still present in the $xy$-orbital of the SrO-terminated system. Indeed, an antiferromagentic  stripe pattern, $\vek{Q}=(0,\pi)$, and symmetrically related at $\vek{Q}=(\pi,0)$, appears, indicated by additional local maxima in the susceptibility, see \fref{fig:suscq}b. Quite notably, in the absence of Hund's rule coupling FM fluctuations are strongly suppressed and said frustrated AF spin-fluctuations would be on par with them (additional data, not shown).
Moreover, we also find incommensurate  magnetic order in the VO$_2$-terminated system around $n=1.3$ in the $xz/yz$-orbitals (iM at $n=1.3$ in \fref{fig:suscq}a). There, instead of a specific ordering vector $\vek{Q}$ the magnetic susceptibility is maximal for all vectors $\vek{Q}$ with origin $(0,0$) and a length of $\delta = \left|\vek{Q}\right|  \geq 0$, i.e., roughly a circle in the $\mathbf{q}$-plane. Upon lowering temperature, $\delta$ increases
and in close vicinity to the ordered phase anisotropy develops; the maximum susceptibility within the circle is found at $\vek{Q}=(\pm\delta,\pm\delta)$, see \fref{fig:incomm}.
These clear maxima suggest that a kind of frustrated ferromagnetism develops where the $xy$-orbital disturbs the alignment of the $xz/yz$-orbitals. Doping beyond $n=1.3$ further increases $\delta$ (data not shown). Let us note here that throughout the phase diagram we did not find any magnetic instabilities supported by Fermi surface nesting.

\medskip

The other prevalent type of instabilities we find are of {\it orbital-ordering} type between the degenerate $xz/yz$ orbitals and can be monitored in the density channel
\begin{eqnarray}
\label{eq:suscoo}
    \chi_d^{xz/yz}(\vek{Q}) & = & \sum_{ij \sigma \sigma'}e^{i\svek{Q}(\svek{R}_i-\svek{R}_j)}\int d\tau \times\\ 
    &\times&\left\langle T_\tau
    (n_{i\, \mathrm{xz}\, \sigma}-n_{i\, \mathrm{yz} \,\sigma})(\tau)(n_{j\, \mathrm{xz}\, \sigma'}-n_{j\, \mathrm{yz}\, \sigma'})(0)
      \right\rangle\nonumber.
\end{eqnarray}

Towards quarter-filling ($n_{i,\sigma}=0.25$) of the $xz$/$yz$-orbitals, i.e.,  at and around stoichiometric filling in the VO$_2$-terminated monolayer and around $n\sim 1.5$ in the SrO-terminated setup, the wave-vector of critical fluctuations is firmly $\vek{Q}=(\pi,\pi)$. As previously alluded to, this leads to a {\it checker-board orbital-order} (cOO), consistent with model expectations \cite{Held1998,PhysRevB.58.R567}.
We note that the $xy$-orbital does not participate in the ordering, as signaled
by susceptibility enhancements being confined to components of the other two orbitals.
The $xy$ orbital can also be passive at larger fillings or valencies:
With one electron more, a t$_{2g}^2$ OO---with $xy$-orbitals near half-filling and one electron alternatingly in the $xz$ and the $yz$-orbital---can occur in YVO$_3$ (LaVO$_3$) if Y (La) ions are partially replaced by Ca (Sr) \cite{PhysRevX.5.011037}, cf.\ Refs.~\onlinecite{Khaliullin2001,Ren2003}.
An OO with all three $t_{2g}$-orbitals participating on the other hand is highly frustrated for a cubic lattice \cite{Khaliullin2000}.

Due to the strong asymmetry around nominal filling in our monolayers, we also find a strong asymmetry of the corresponding cOO-dome in \fref{fig:phasediagram}a, where the cOO transition temperature even increases upon electron-doping. If we move too far away from ideal quarter-filling, the ordering temperature is suppressed rapidly.
Despite this suppression of cOO we find an additional emerging ordering for $n\sim0.9$ in \fref{fig:phasediagram}a. The corresponding ordering can again be described via Eq.~\eqref{eq:suscoo}
with, however, a characteristic vector
$\vek{Q}=(0,\pi)$ (and $\vek{Q}=(\pi,0)$ related via symmetry), describing {\it stripe orbital-ordering} (sOO).
The cOO-to-sOO transition under hole-doping is not realized by a continuous move of the ordering vector from $\vek{Q}=(\pi,\pi)$ to $\vek{Q}=(0,\pi)$.
Instead, increased hole-doping suppresses cOO while simultaneously promoting sOO.
We conjecture that this transition can be ascribed to the `auxiliary' $xy$-orbital,
that does not contribute to the susceptibility enhancements of either fluctuations.
Illustrated in the inset of \fref{fig:occupation} we find a semi-empirical condition
that links the preference for stripe over checkerboard orbital order to the filling of the $xy$-orbital:
$n_{\mathrm{xy},\sigma} = (0.25-n_{\mathrm{xz/yz},\sigma})/2)$ below which checkerboard ordering and above which stripe ordering is preferred by the system. Effectively, enough $d_{xy}$ occupation frustrates the local site enough for stripe ordering to be energetically favorable. As for the electron-doped side, the orbital-ordering domes in both systems disappear when doping too far away from quarter-filling.

\section{Discussion}
\label{Sec:Discussion}

Mapping out all these different magnetic and orbital-ordering instabilities and their corresponding critical temperatures yields the phase diagrams in \fref{fig:phasediagram}.
Naturally, the DMFT long-range order exhibits mean-field criticality (Gaussian fluctuations), i.e., the critical exponents are $\gamma=1$ for the susceptibility $\chi$, and $\nu=0.5=\gamma/2$ for the correlation length $\xi$, see \fref{fig:suscq}.
We note that in strictly two dimensions, due to the Mermin-Wagner theorem \cite{Mermin1966}, long-range order can only set in at $T=0$. As a spatial mean-field theory, DMFT does not verify this constraint, while it is captured in, e.g., D$\Gamma$A\cite{Rohringer2011,Schaefer2020}.
In an experimental setting, however, perturbations by disorder (oxygen vacancies, etc.), spin-orbit coupling, single-ion anisotropy, surface adatoms, as well as tunneling into the substrate render strict 2Dness obsolete. This allows for phase transitions at finite temperatures.
Still, non-local correlations beyond DMFT\cite{RevModPhys.90.025003,Tomczak2017review} are expected to attenuate the tendency for long-range order with respect to our DMFT solution.
Non-local fluctuations may even drive pseudogaps, as shown for antiferromagnetic fluctuations in the one-band Hubbard model\cite{Schaefer2015-2,Schaefer2015-3,Gukelberger2016,Schaefer2020}.

In \fref{fig:phasediagram} we further indicate that instead of chemical doping, the phase diagram can be perused by applying a gate voltage $V_G$. We note that the estimated values of $V_G$ measure the necessary potential directly in the monolayer, not the truly external one applied at a certain distance through a dielectric medium. Realizing the whole phase diagram with a sheet carrier change of 0.5
electrons/unit cell ($3\cdot10^{14}$ electrons/cm$^2$) in a single device
might be challenging\cite{RevModPhys.78.1185}. Ionic electrolytes as dielectrics, however, allow for such a large amount of induced charges thanks to the large capacitance
of polarized ions\cite{Katase3979}. Experimentally, the phase diagram \fref{fig:phasediagram} can hence be realized either by doping or via  gate voltage, or a combination of both.

Finally, let us comment on an important difference between our semi-{\it ab initio} calculations and typical model setups.
In multi-orbital (Hubbard) {\it models}, orbital complexity is often simplified, in the sense that the hopping of each orbital is considered to be {\it isotropic} vis-à-vis all neighboring atoms. Many model studies even employ semi-circular densities of states, corresponding to the Bethe lattice with infinite coordination number.
Such simplifications allowed distilling essential behaviors
of, e.g., correlation enhancements of crystal-fields\cite{PhysRevB.78.045115},
the influence of Hund's physics on the Mott transition \cite{PhysRevLett.107.256401},  or correlations in band-insulators \cite{PhysRevB.80.155116,NGCS}.
Indeed, qualitative {\it spectral properties} of 3D systems without broken symmetries,
are mainly controlled
by the orbitals' filling and the kinetic energy they mediate.
Near symmetry-broken phases, however, the associated fluctuations
in {\it physical susceptibilities} are strongly dependent on electronic-structure details. In fact, van-Hove singularities, nesting, or Kohn points may well be the cause of an instability.
In our context, besides anisotropies, e.g., induced by the tetragonal distortion\cite{Benjamin_2D3D}
and the obvious geometric restriction,
a crucial ingredient is the {\it per se} 2D nature of the $t_{2g}$-orbitals:
In perfectly 3D cubic perovskites (e.g., bulk SrVO$_3$) each transition-metal $t_{2g}$ orbital only hybridizes with four of the six oxygen ions of the coordination octahedron.
Therefore, both in bulk and ultra thin films, the in-plane $d_{xy}$ DOS is 2D-like.
The loss of hopping along the $z$-direction in ultra-thin films
further reduces the effective dimensionality, leading to the 1D-like DOS of the $d_{xz,yz}$ orbitals, see \fref{Fig:dft}. %
Our analysis suggests, that even for qualitative phase diagrams of ultra-thin oxide films or heterostructures,  both, the crystal geometry and the orbital structure have to be accounted for.

\section{Conclusion and outlook}
\label{Sec:Conclusion}

\begin{figure}[!t]
  \includegraphics[width=0.4\textwidth]{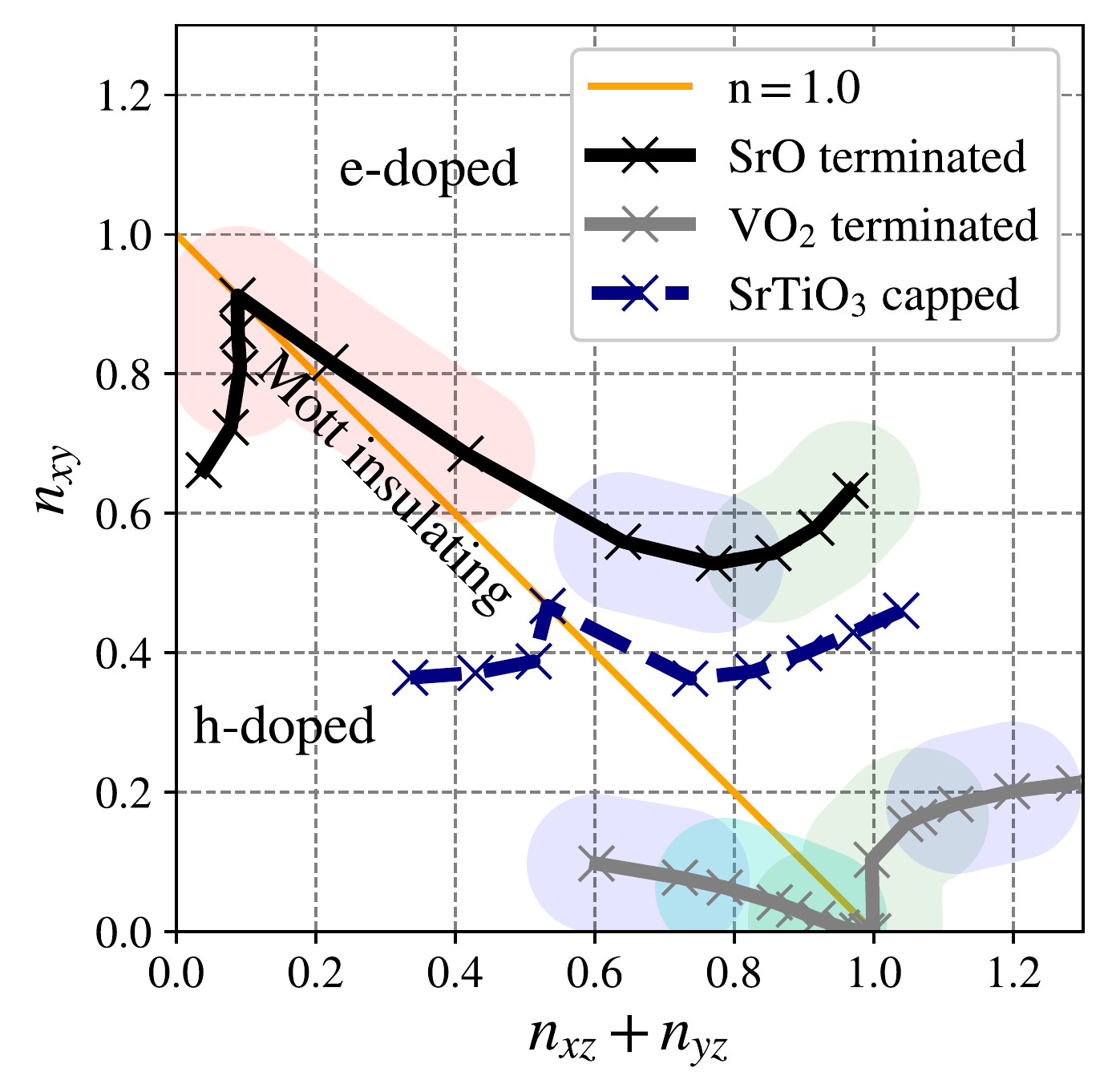}
  \caption{{\bf Global phase diagram of SrVO$_3$ monolayers on SrTiO$_3$.} DMFT orbital occupations of each setup are mapped into a $n_{xy}$ vs.\ $n_{xz} + n_{yz}$ graph. Due to the reduced bandwidth in the monolayer on a SrTiO$_3$ substrate, any orbital occupation repartitioning 
  of the nominal $n=1$ filling (orange line) realizes a Mott insulator. SrO-terminated (black) and VO$_2$-terminated (gray) monolayers and their dominant non-local fluctuations (shaded background; colors identical to \fref{fig:phasediagram}, but lighter) are discussed in this publication. As an outlook we showcase the effect of embedding a SrVO$_3$ monolayer in a SrTiO$_3$ sandwich (navy blue, dashed): The crystal-field and orbital structure is somewhat intermediate to the two uncapped monolayers. The computation of respective instabilities is left for future work.
  }
  \label{fig:orbmap}
\end{figure}

We have studied a single layer of SrVO$_3$ on a SrTiO$_3$ substrate, using DFT and DMFT, considering both possible surface terminations, VO$_2$ and SrO, to vacuum.
We demonstrated that stoichiometric samples ($n=1$) are
bandwidth-controlled Mott insulators:
Depending on the surface termination, SrO or VO$_2$, the monolayer is an effectively half-filled one-orbital or a quarter-filled two-orbital Mott insulator.
We showed this orbital polarization to derive from the crystal-field splitting
having opposite signs for the two terminations and to be significantly enhanced by electronic correlations.
Electron or hole-doping reveals multi-orbital effects:
For the SrO-termination, AF-fluctuation are dominant around nominal filling. Doping with electrons populates the  $xz/yz$-orbitals; they order ferromagnetically  ($n\sim1.3$) or realize checkerboard orbital orbital order ($n\sim1.5$).
For the VO$_2$ termination checkerboard $xz/yz$ orbital-order already dominates around nominal filling. Doping then instead promotes the $xy$-orbital which acts as a mediator for ferromagnetism and stripe orbital-order on the hole-doped side and incommensurate magnetism
on the electron-doped side.
While the change in magnetic fluctuations and orders could be observed in neutron experiments, experimentally evidencing the orbital fluctuations is only possible indirectly: the staggered pattern of $xz$ and $yz$-orbitals will result in a dynamic (potentially static) alternation of the bond-length in the $x$ and $y$ direction, possibly detectable in future x-ray measurements.

In all, the orbital polarization is the essential driver of the phase diagram of the SrVO$_3$ monolayer on SrTiO$_3$.
We therefore summarize our results in Fig.~\ref{fig:orbmap} in form of
an orbital occupation map.
The considered surface terminations each realize, under doping, a characteristic trajectory in the $n_{xy}$ vs.\ $n_{xz}+n_{yz}$ space.
As an outlook, we include a third possibility---a SrVO$_3$ monolayer with SrTiO$_3$ on both sides: At nominal filling this sandwich is, again, a Mott insulator. However, owing to the symmetric embedding, the crystal-field is
minute (but positive).
The computation of ordering instabilities of capped ultra-thin films, in which quantum confinement effects could be studied in a more controlled fashion, is left for future work.

A different future avenue are even more realistic setups of the current geometries, e.g., atomic position relaxation with the inclusion of correlation effects and recent advances\cite{PhysRevLett.112.146401, PhysRevB.102.245104} which allow for calculations of forces and phonons within DFT+DMFT to test the dynamical stability against superstructure formations.

\begin{acknowledgments}
  We thank  R. Claessen, M.\ Fuchs, J. Gabel, A.\ Galler,  G.\ Sangiovanni, M. Sing, P.\ Thunstr\"om and Z.\ Zhong  for fruitful discussions.
 The authors acknowledge support from the Austrian Science Fund (FWF) through grants P 30819, P 30997, P 32044, and P 30213. Calculations were performed in part on the Vienna Scientific Cluster (VSC).
\end{acknowledgments}

\end{document}